\documentclass[a4paper]{article}
\usepackage{amsmath,amstext,amsgen,amsbsy,amsopn,amsfonts,amssymb}
\usepackage{easybmat}
\usepackage{graphics}
\usepackage[pdftex]{graphicx}
\usepackage{epsfig}
\usepackage{amsthm}
\usepackage{bm}
\usepackage{algorithm}
\usepackage{algorithmic}

\usepackage{wrapfig}
\usepackage{esvect}
\usepackage{multirow}
\usepackage{array}
\usepackage{fancyvrb}
\usepackage{gensymb}

\begin{document}
\title{\textbf{The Turtleback Diagram for \\Conditional Probability}}
\author{
Donghui Yan and Gary E. Davis
\vspace{0.1in}\\
Department of Mathematics and Program in Data Science\vspace{0.05in}\\
University of Massachusetts Dartmouth, MA\\[0.05in]
}

\date{}
\maketitle

\begin{abstract}
We elaborate on an alternative representation of conditional probability to the usual tree diagram. We term the representation ``turtleback diagram'' for its resemblance to the pattern on turtle shells. Adopting the set theoretic view of events and the sample space, the turtleback diagram uses elements from Venn diagrams---set intersection, complement and partition---for conditioning, with the additional notion that the area of a set indicates probability whereas the ratio of areas 
for conditional probability. Once parts of the diagram are drawn and properly labeled, the calculation of conditional probability involves only simple arithmetic on the area of relevant sets. We discuss turtleback diagrams in relation to other visual representations of conditional probability, and detail several scenarios in which turtleback diagrams prove useful. By the equivalence of recursive space partition and the tree, the turtleback diagram is seen to be equally expressive as the tree diagram for representing abstract concepts. We also provide empirical data on the use of turtleback diagrams with undergraduate students in elementary statistics or probability courses.
\end{abstract}
\section{Introduction}
Conditional probability \cite{Rice1995, JohnsonTsui2003,Ancker2006, Mann2010} is an important
concept in probability and statistics. It has been widely acknowledged that the concept of conditional probability,
and particularly its application in practical contexts, is  difficult for students \cite{TverskyKahneman1980, FischbeinGazit1984,Falk1986, TarrJones1997,TomlinsonQuinn1997, Tarr2002, Yanez2002,TarrLannin2005} and especially those without much background or
previous training in mathematics at the college level. 
\\
\\
Let $A$ and $B$ be two events, then the conditional
probability of $A$ given $B$ is defined as
\begin{equation}
\label{eq:condDef}
\mathbb{P}(A|B)=\frac{\mathbb{P}(A \cap B)}{\mathbb{P}(B)}.
\end{equation}
Our experience with undergraduate students is that a major difficulty in understanding and working effectively with conditional probability lies in the level of abstraction involved in the concepts of ``event'' and ``conditioning''; see also \cite{Falk1986}.
\\
\\
The focus of this article is on productive visual representations for the understanding and application of conditional probability. The significant role of visual representation in mathematics is well-established; see, for example, \cite{Arcavi2003, Presmeg2006}. While visualization is an important topic in statistics (see, e.g., \cite{Tukey1977, Cleveland1993}), the role of visualization in statistics education or practice is not as well documented. In particular, there is actually not much research into productive visualization of conditional probability \cite{Morris2016,Collins2017}; popular 
books such as \cite{GelmanCarlin2013} do not dedicate much effort to visual explanations of the Bayes theorem. There has been 
some research on school student difficulties with conditional probability \cite{FischbeinGazit1984, TarrJones1997, Tarr2002, Yanez2002, TarrLannin2005} but much less so for undergraduates. Our aim in discussing turtleback diagrams is to provide a visual tool for the representation of conditional probability that may, additionally, be used in further research on student understanding of conditional probability.

\section{Student difficulties in understanding conditional probability}
Tomlinson and Quinn \cite{TomlinsonQuinn1997}, in discussing their graphic model for representing conditional probability (see Section~\ref{Tomlinson-Quinn}), state:
\begin{quote}
``Conditional probability is a difficult topic for students to master. Often counter-intuitive, its central laws are composed of abstract terms and complex equations that do not immediately mesh with subjective intuitions of experience. If students are to acquire the mathematical skills necessary for rational judgement, teaching must focus on challenging the personal biases and cognitive heuristics identified by psychologists, and demonstrate in the most accessible way---the power of probabilistic reasoning.'' (p.7) 
\end{quote}
Documented student difficulties with conditional probability can be summarized as one of three main types \cite{Falk1986}:
\begin{enumerate}
\item Interpreting conditionality as causality.
\item Identifying and describing the conditioning event.
\item Confusing $\mathbb{P}(A\vert B)$ and $\mathbb{P}(B\vert A)$.
\end{enumerate}
Tarr and Jones \cite{TarrJones1997} developed a valid and reliable framework for addressing student difficulties with conditional 
probability, in the context of sampling without replacement. This framework is particularly valuable in carrying out research as to 
which visual representation of conditional probability is most useful in assisting students and teachers. 
\section{Visual representations of conditional probability}
\subsection{Tree diagrams}
\begin{figure}[ht]
\centering
\begin{center}
\hspace{0cm}
\includegraphics[scale=0.32,clip]{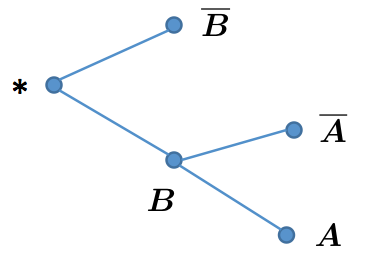}
\end{center}
\abovecaptionskip=-5pt
\caption{\it The tree diagram approach for conditional probability. } \label{figure:treeDiagram}
\end{figure}

Tree diagrams have been used by many to help understand conditional probability. The idea of a tree diagram is to use nodes for 
events, the splitting of a node for sub-events, and the edges in the tree for conditioning. For example, Figure~\ref{figure:treeDiagram} 
is an illustration of conditional probability. Node $*$ indicates the sample space $\Omega$, and we will use them interchangeably 
throughout. Two possible events, either $B$ or $\overline{B}$, may happen. This is represented by two tree nodes $B$ and $\overline{B}$. 
The splitting of node $B$ into two nodes $A$ and $\overline{A}$ indicates that, given $B$, two possible events, $A$ and $\overline{A}$, 
may occur. The edges, $B \rightarrow A$ and $B \rightarrow \overline{A}$ indicate conditional probabilities, $\mathbb{P}(A | B)$ and 
$\mathbb{P}(\overline{A} | B)$, respectively. 
Tree diagrams help many students to understand the concept of conditional probability and apply it for problem solving, 
but is not so effective to many others especially those less prepared ones. Basically, they find the following two aspects 
non-intuitive. One is to represent events by tree nodes, which usually appear as dots or small circles, but events are sets 
and are more naturally represented by Venn diagram \cite{Edwards2004} type of notations. Another is the idea to represent 
conditional probability by tree edges; it is hard to see any straightforward connections of this to \eqref{eq:condDef}.
\\
\\
To address issues with the tree diagram, let us re-examine the idea of graphical visualization. There are two important 
ingredients (or steps) in visualizing an abstract mathematical concept. One is {\it a concrete graphical representation} 
of the target mathematical objects. This step would offload part of the burden of the brain by concrete graph objects, 
without which one has to keep relevant abstract mathematical objects in the brain and gets ready for subsequent 
mathematical operation. The second is that, the mathematical concept or operation can be understood or achieved by 
{\it a simple operation on the graphical objects}. This is the step to be carried out in the brain, and preferred to be simple 
(or at least conceptually simple). If a balance could be achieved between these two ingredients in visualizing a mathematical 
concept, then the graphical tool would be successful. This explains why the {\it Venn diagram} has been so successful 
since it was introduced, and has now become the standard graphical tool for set theory. Essentially, the Venn diagram 
converts the set objects to graph objects in such a way that many set relationships or operations could be accomplished 
by `reading' the diagram---the mathematical operation is done directly by the human visual system, instead of having to 
invoke both the visual system and the brain. On the other hand, for the tree diagram, each of the two ingredients does 
some job but there is room for improvement. 
\\
\\
The turtleback diagram we propose tries to optimize the two steps involved in the design of a graphical tool for conditional 
probability. In particular, it views events and the sample spaces as sets, and uses elements from Venn diagrams---set intersection, 
complement and partition---for conditioning, with the additional notion that the area of a set indicates probability whereas 
the ratio of areas associated with relevant sets indicates conditional probability. Once parts of the diagram are drawn and 
properly labelled, the calculation of conditional probability involves just simple arithmetic on the area of relevant sets. This 
makes it particularly easy to understand and use for problem solving.

\subsection{Other visual representations}
There have been several prior attempts to represent conditional probability visually \cite{GigerenzerHoffrage1995,TomlinsonQuinn1997, 
Yamagishi2003, Brase2009}, and we discuss briefly three of these below.
\subsubsection{Tomlinson-Quinn graphical model}
\label{Tomlinson-Quinn} 
This graphical model, for facilitating a visually moderated understanding of conditional probability, described in \cite{TomlinsonQuinn1997}, 
is a modified tree diagram. 
\\
\\
Tomlinson and Quinn visualize compound events $A \cap B, A \cap \overline{B}$ as nodes of a tree (see Figure 2 of \cite{TomlinsonQuinn1997}, 
so essentially their idea is still a tree diagram in which they carry out a Venn-diagram like visualization at each tree node. 
\subsubsection{Roullete-wheel diagrams} 
Yamagishi \cite{Yamagishi2003} introduces roullete-wheel diagrams as a visual representation tool; see Fig 1, p. 98 of \cite{Yamagishi2003}. 
He argues that 
\begin{quote}
``the graphical nature of [\textit{roulette-wheel diagrams}] take advantage of people's automatic visual computation in grasping the relationship 
between the prior and posterior probabilities.'' (p. 105).
\end{quote}
and provides experimental evidence that use of roulette-wheel diagrams increases understanding of conditional probability beyond that for tree 
diagrams. In this regard, Sloman et al. \cite{Sloman2003} state:
 \begin{quote}
 ``The studies reported support the nested-sets hypothesis over the natural frequency hypothesis. \ldots The nested-sets hypothesis is the 
 general claim that making nested-set relations transparent will increase the coherence of probability judgment.'' (p. 307)
 \end{quote}
\subsubsection{Iconic diagrams}
\begin{figure}[ht]
\centering
\begin{center}
\hspace{0cm}
\includegraphics[scale=0.32,clip]{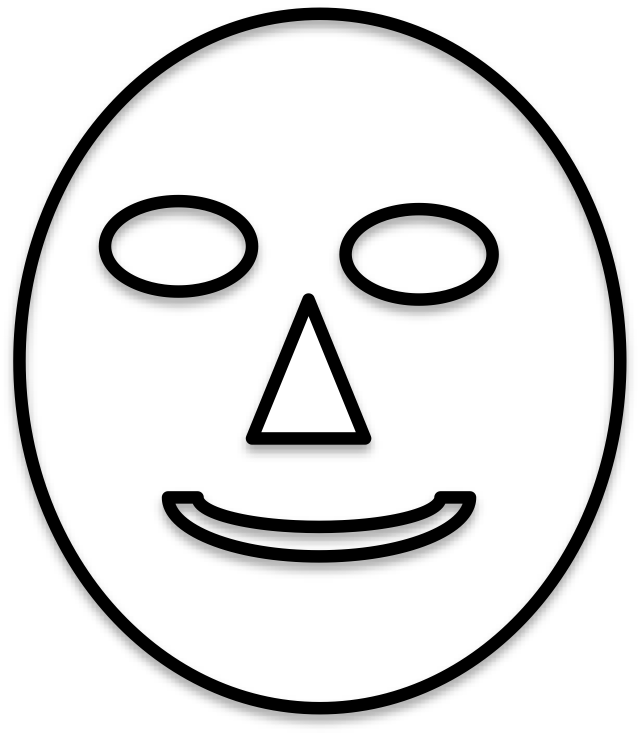}
\end{center}
\abovecaptionskip=-2pt
\caption{\it A diagram that is universally iconic for humans. } \label{figure:face}
\end{figure}

\begin{figure}[ht]
\centering
\begin{center}
\hspace{0cm}
\includegraphics[scale=0.4,clip]{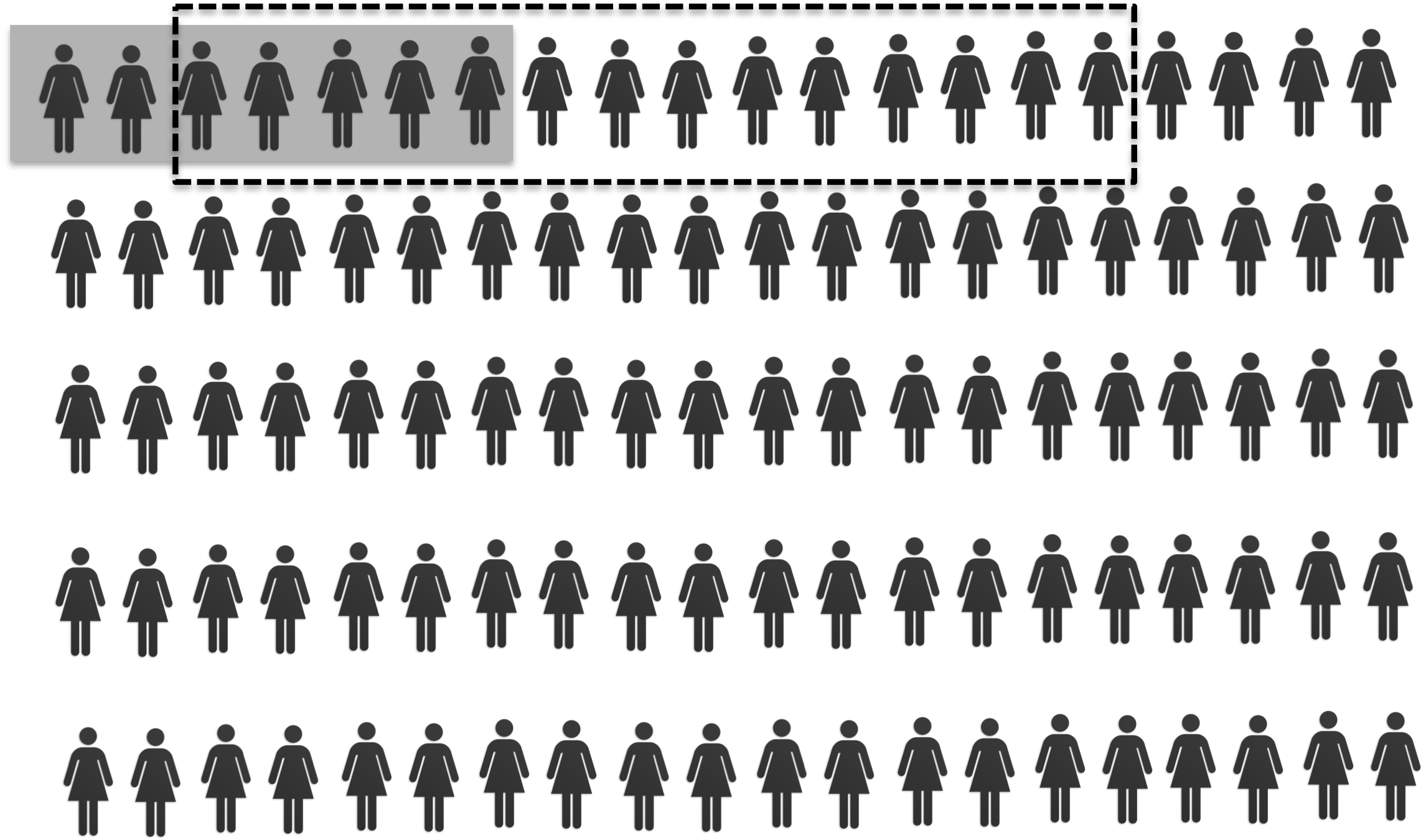}
\end{center}
\abovecaptionskip=-2pt
\caption{\it An iconic representation of the effectiveness of a cancer test.} \label{figure:faceCancer}
\end{figure}

``Iconicity'' is the lowest of Terrence Deacon's three levels of symbolic interpretation\footnote{In Deacon's framework, there are three levels 
of referential relationship in a cognitive process, including iconic, indexical, and symbolic reference, where higher levels are built hierarchically 
upon lower levels.} \cite{Deacon1998}, as it is for Peirce on whose semiotic work Deacon's theory is based. An \textit{icon} is a form of graphical 
representation that requires no significant depth of interpretation: an icon brings to mind, without any apparent intermediate thought, something 
that it resembles. For example, the diagram in Figure~\ref{figure:face} is universally iconic for human beings.
Brase \cite{Brase2009} carried out a number of experiments from which he inferred that an iconic representation of a Bayesian probability question is more effective in eliciting correct responses than either no visual aids, or Venn diagrams.
\\
\\
A modified version of Brase's question is as follows:
\begin{quote}
``A new test has been developed for a particular form of cancer found only in women. This new test is not completely accurate. Data from other tests indicate a woman has 7 chances out of 100 of having cancer. The test indicated positively only 5 of these women as having cancer. On the other hand, the test indicated a positive result for 14 of the 93 women without cancer. 
\\\\
Janine is tested for cancer with this new test. Janine has probability \rule{0.5cm}{0.01cm} of a positive result from the test, with a probability \rule{0.5cm}{0.01cm} of actually having cancer."
\end{quote}
An iconic representation for this problem is shown in Figure~\ref{figure:faceCancer}.
\\
The strength of such iconic representations is that they reduce the calculation of probabilities to simple counting problems and, as Brase \cite{Brase2009} demonstrates, are effective in assisting students to get correct answers. A weakness of iconic representations such as these, are that they rely on counting discrete items and so are quite limited in representing more realistic probabilities. 
\subsection{Turtleback diagrams}
\label{section:spacePartition} 
Our focus is on how to represent an event graphically, how to relate it to the sample space, how to express the notion of conditioning such that it would be easy to understand the concept of conditional probability, to gather pieces of information together, and to solve problems accordingly.
\\
\\
We start by treating the sample space (denoted by $\Omega$) and events as sets, 
and in terms of graph, as a region and its sub-regions, similarly as in a Venn diagram. Assume the region
representing the original sample space $\Omega$ has an area of $1$.
To simplify our discussion (or to abuse the notation), we will use a label,
say $``B"$, to denote the region associated with event $B$. Note that here the
label can be either a single letter, or several letters ({\it such a case indicates
the intersection of events}. For example, a label $``AB"$ indicates the intersection
of events $A$ and $B$ and thus that of regions $A$ and $B$). Similarly we can
use the union of two regions (viewed as sets) to represent the union of two
events. Other operations of events can also be defined accordingly in terms of set operations; we omit
the details here. To quantify the chance of an event, we associate it with the
area of the relevant region. For example, $\mathbb{P}(B)$ is indicated by the
area of region $B$.
\\
\\
The centerpiece in `graphing' conditional probability is to express the notion of conditioning. 
This can be achieved by re-examining the definition of conditional probability as given in \eqref{eq:condDef}. 
It can be interpreted as follows. Let $A$ be the event of interest. Upon conditioning, say, on 
event $B$, both the new effective sample space and event $A$ in this new sample space can 
be viewed as their restriction on $B$, that is, $\Omega$ becomes $\Omega \cap B=B$ and 
$A$ becomes $A \cap B$, respectively. The conditional probability $\mathbb{P}(A|B)$ can 
now be interpreted as the proportion of the part of $A$ that is inside $B$ (i.e., $A \cap B$) 
out of region $B$, that is,
\begin{equation}
\mathbb{P}(A|B)=\frac{\mbox{area of region~} A \cap B}{\mbox{area of~}B}.
\end{equation}
Now we can describe how to sketch a turtleback diagram. We start by drawing a circular disk which represents the sample 
space $\Omega$. Then we represent events by partitioning the circular disk and the resulting subregions. To facilitate our 
discussion, we define the partition of a set \cite{ChartrandZhang2011}. $\mathcal{P}=\{S_i|~i \in \mathbb{I}\}$ is a 
partition of set $S$ if $S=\cup_{i} S_i$ and $S_i \subseteq S$, $S_i \cap S_j=\emptyset$ for all $i \neq j \in \mathbb{I}$. 
\begin{figure}[ht]
\centering
\begin{center}
\hspace{-0.1in}
\includegraphics[scale=0.32,clip]{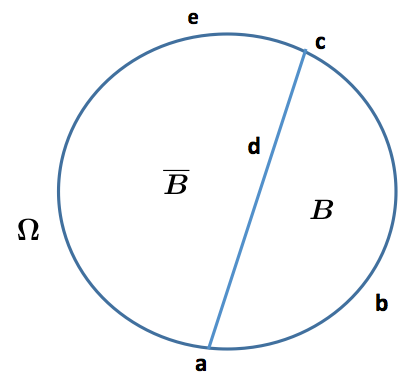}
\hspace{0.1in}
\includegraphics[scale=0.32,clip]{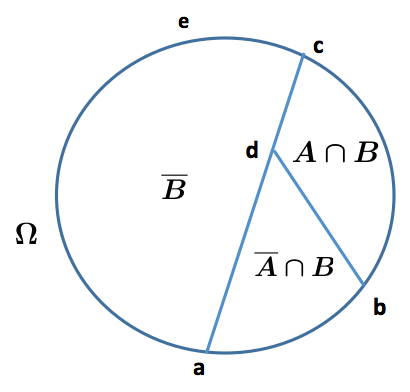}
~~\quad
\includegraphics[scale=0.32,clip]{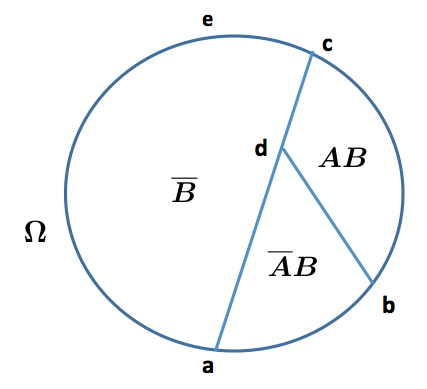}
\end{center}
\abovecaptionskip=-5pt
\caption{\it Illustration of the turtleback diagram for conditional probability. 
The left panel shows the partition of $\Omega$ by $\Omega=B \cup \overline{B}$, the middle panel shows event 
$B$ is further partitioned by $B=(A \cap B) \cup (\overline{A} \cap B)$, and the right panel is a simplified version of
the middle panel where $``AB"$ stands for $A\cap B$, and $``\overline{A}B"$ stands for $\overline{A} \cap B$. The 
conditional probability $\mathbb{P}(A | B)$ is the ratio of the area for region ``bcdb" and that for the area ``abcda".
} \label{figure:condProbDef}
\end{figure}
\\
\\
We will use Figure~\ref{figure:condProbDef} to assist our description.
To represent the partition $\Omega = \overline{B} \cup B$, we use a {\it straight line} ``adc" to split the circular disk into two halves, i.e.,
regions surrounded by ``abcda" and ``adcea", which stands for event $B$ and $\overline{B}$, respectively. The regions corresponding
to event $B$ and $\overline{B}$ can be further split for a more refined representation involving other events. To represent conditional
probability as defined by \eqref{eq:condDef}, event $B$ is written as 
\begin{equation}
\label{eq:totalProbFormula}
B= (A \cap B) \cup (\overline{A} \cap B),
\end{equation}
which can be represented by splitting the region for $B$, i.e., ``abcda", with a straight line ``db". The conditional 
probability $\mathbb{P}(A | B)$ can then be calculated as the ratio of the area for region ``bcdb" and that for region
``abcda". 
\\
\\
The turtleback diagram leads to a partition of the sample space $\Omega$ as follows
\begin{eqnarray}
\label{eq:recurPar1} \Omega &=& \overline{B} \cup B\\
&=&
\label{eq:recurPart2} \overline{B} \cup (A \cap B) \cup (\overline{A} \cap B).
\end{eqnarray}
Continuing this process, we can define
events as complicated as we like in a simple hierarchical
fashion as a nesting sequence of partitions $\mathcal{P}_0 \succ \mathcal{P}_1 \succ \mathcal{P}_2 \succ ...$ 
where $\mathcal{P}_0=\{\Omega\}$, $\mathcal{P}_1=\{B, \overline{B}\}$, and $\mathcal{P}_{i+1}$ is a refinement
of $\mathcal{P}_{i}$ for index $i>0$ in the sense that each element
in $\mathcal{P}_{i+1}$ is a subset of some element in $\mathcal{P}_{i}$.
\\
\\
We can now {\it assign labels to each of the sub-regions}, e.g., by the name of
the relevant events to indicate that a particular region is associated with that event. 
For example in Figure~\ref{figure:condProbDef}, we assign labels $``AB"$ and
$``\overline{A}B"$ to regions ``bcdb" and ``abda", respectively. Here, $``AB"$ means $A\cap B$, 
and $``\overline{A}B"$ indicates $\overline{A} \cap B$, and the same convention carries over throughout. 
Accordingly, the turtleback diagram simplifies 
to the right panel in Figure~\ref{figure:condProbDef}. Note that here
an event need not be a connected region, rather it could be a collection of
patches (i.e., small regions) with each of them capturing
information from a different source. This causes a little burden in
calculation but costs really nothing {\it conceptually}, or, in terms of the ability of
visualization.
\\
\\
One advantage of such a recursive-partition representation of the sample space
$\Omega$ is that the data are now highly organized and we can easily
operate on it, for example to find out the probability of a certain event.
The idea of organizing the data via recursive space-partition and manipulating by
their labels has been explored in CART (classification and regression trees \cite{CART}) 
and more recently, random projection trees \cite{RPTree}, as
well as a recent work of one author and his colleagues \cite{YanHJ2009}. Note that dividing a
region into a number of small patches also entails the total probability
formula, an important ingredient in conditional probability to which formula
\eqref{eq:totalProbFormula} is related. We will use the `Lung disease and
smoking' example to illustrate the use of turtleback diagrams for conditional probability.
\begin{figure}[h]
\centering
\begin{center}
\hspace{0cm}
\includegraphics[scale=0.32,clip]{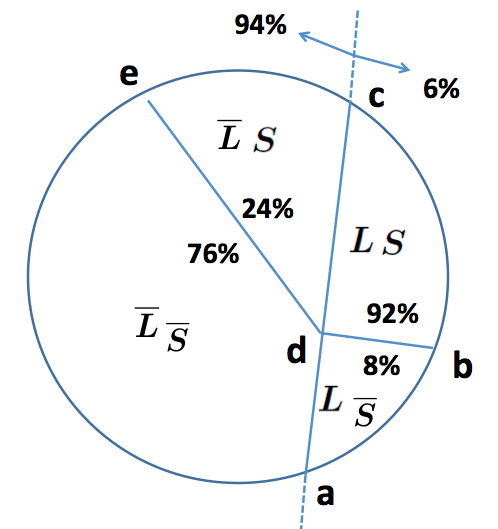}
\end{center}
\caption{\it The turtleback diagram for the `Lung disease and smoking'
example. The letters $``L"$ and $``\overline{L}"$ stand for ``with lung disease" and
``without lung disease", $``S"$ and ``$\overline{S}$" for ``smoking" and ``nonsmoking",
respectively. } \label{figure:LSg}
\end{figure}
\subsection{The lung disease and smoking example}
\label{section:lingSmokerg} This example is taken from online sources (see \cite{Weiss2012}).  
It is described as follows.
\begin{quote}
``According to the Arizona Chapter of the American Lung Association, $6.0\%$
of population have lung disease. Of those having lung disease, $92.0\%$ are
smokers; of those not having lung disease, only $24.0\%$ are smokers. Answer
the following questions.
\begin{enumerate}
\item[(1)]  If a person is randomly selected in the population, what is
    the chance that she is a smoker having lung disease?
\item[(2)]  If a person is randomly selected in the population, what is
    the chance that she is a smoker?
\item[(3)] If a person is randomly selected and is discovered to be a
    smoker, what is the chance that she has lung disease?"
\end{enumerate}
\end{quote}
According to the information given in the problem, we can sketch a graph as
Figure~\ref{figure:LSg}. Labels and area information to each sub-regions are
assigned properly. Assume the circular disk has an area of $1$.
Now we can answer the questions quickly as follows.
\begin{enumerate}
\setlength\itemsep{-0.02in}
\item[(1)] The answer is simply the area of region ``adba", which is $6\%
    \cdot 92\%=0.0552$.
\item[(2)] The answer is the area of region ``edbae", which is
    $6\% \cdot 92\%+94\% \cdot 24\%=0.2808$. This is, in essence, the total
    probability formula $\mathbb{P}(S)=\mathbb{P}(L \cap
    S)+\mathbb{P}(\overline{L} \cap S)$.
\item[(3)] Recognizing that this involves conditional probability and is the ratio of two relevant areas,\\
    (area of ``adba"/area of
    ``edbae")=0.0552/0.2808=0.1966.
\end{enumerate}
\subsection{Difficulty with the Venn diagram}
\begin{figure}[h]
\centering
\begin{center}
\hspace{-0.1in}
\includegraphics[scale=0.4,clip]{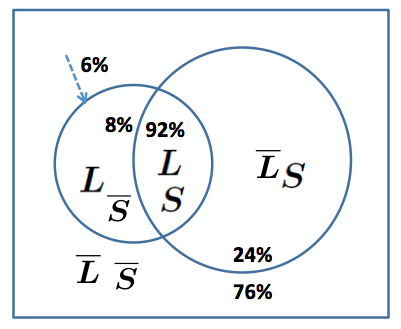}
\end{center}
\caption{\it The Venn diagram approach to the `Lung disease and smoking' example.
} \label{figure:LSVenn}
\end{figure}

The Venn diagram is known as the standard graphical tool for set theory. Both Venn diagram and the 
turtleback diagram use regions to represent sets. However, there is a major difference. In a turtleback 
diagram, as illustrated in Figure~\ref{figure:condProbDef}, {\it straight lines}, such as line ``adc", ``db" etc, 
are used to split the sample space and regions. In contrast, the Venn diagram represents events by 
drawing {\it circular disks}. Partitioning the sample space $\Omega$ in such a way would cause substantial 
difficulty in handling the complement operation, one crucial ingredient in conditional probability. One has 
to deal with a setting where the complement of a region would surround the region itself, for example, in 
Figure \ref{figure:LSVenn}, $``\overline{S}"$ and $``\overline{L}"$ surround $``S"$ and $``L"$, respectively. This 
would cause extra burden to the human brain or the visual system. We will illustrate with the `Lung disease 
and smoking' example.
\\
In Figure~\ref{figure:LSVenn}, one would find it
tricky to label the region and put area information for $``\overline{L}"$ (which is $94\%$) 
without causing confusion. Moreover, it may require some extra work (versus simply
``reading" from the graph) to assign
the label $``\overline{L}\overline{S}"$, or to calculate the area of this region. In contrast,
the turtleback diagram (c.f. Figure~\ref{figure:LSg}) introduces {\it straight lines}, 
e.g., ``adc", ``ed", and ``db", which readily avoids obstacles caused by set intersections
or complements in a Venn diagram. 
\section{Semantic equivalence of the turtleback and the tree diagram}
\label{section:treeBased} 
\begin{figure}[ht]
\centering
\begin{center}
\hspace{0cm}
\hspace{-0.25in}\includegraphics[scale=0.32,clip]{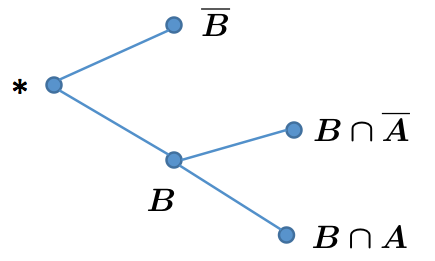}
\hspace{0.2in}
\includegraphics[scale=0.32,clip]{part2Trees2.png}
\end{center}
\caption{\it The tree diagram representation of
the turtleback diagram in Figure~\ref{figure:condProbDef}. } \label{figure:part2Tree}
\end{figure}

Given a graphical representation, it is natural to ask questions about its expressive
power--will it be expressive enough to represent a complicated or very abstract concept? 
We will show that the turtleback diagram is equally expressive as the tree diagram.
\\
\\
The way that the turtleback diagram progressively 
refines the partition over the sample space is essentially a recursive space
partition, where the sets involved in the
partition are organized as a chain of enclosing sets. For example, in
Figure~\ref{figure:condProbDef}, we have
\begin{equation}
\left(A \cap B\right) \subseteq B \subseteq \Omega,
~\mbox{and}~\left(\overline{A} \cap B\right) \subseteq B \subseteq
\Omega.
\end{equation}
By equivalence (see, for example, \cite{CART}) between the
recursive space partition and the tree structure, we can actually show the ``semantic" equivalence 
between the turtleback diagram and the tree diagram. The remaining of this section is dedicated to this.
Let a tree node correspond to a set in a recursive space partition with the following three properties:
\begin{itemize}
\setlength\itemsep{-0.02in}
\item[1)] The root node corresponds to the
sample space $\Omega$; 
\item[2)] All the child nodes of a node form a
decomposition of this node; 
\item[3)] Down from the root node, the nodes along
any path form a chain of enclosing sets. 
\end{itemize}
Property 2) entails the total probability formula, and property 3) corresponds to a refinement of
a partition. This allows one to turn the turtleback diagram
in Figure~\ref{figure:condProbDef} into a tree representation, that is, 
the left panel of Figure~\ref{figure:part2Tree}. The ``chain" property forces a child node
to be a restriction of its parent node. We can use this to simplify the labels for the
tree nodes, e.g., the left panel becomes the right in Figure~\ref{figure:part2Tree}. 
Note that in the right panel, really node $A$
corresponds to the set $\Omega \cap B \cap A$, that is, the intersection
of all sets along the path from the root to node $A$ (i.e., the tree
path $* \rightarrow B\rightarrow A$). 
\\
\\
For real world conditional probability problems, often the following formula is
used instead of \eqref{eq:condDef}, due to availability of information from multiple sources
\begin{equation}
\label{eq:condDef2}
\mathbb{P}(A|B)=\frac{\mathbb{P}(A \cap B)}{\sum_i\mathbb{P}(B \cap A_i)}
\end{equation}
where $\sum_i A_i=\Omega$. This requires the calculation of probabilities in the form
of $\mathbb{P}(B \cap A_i)$, or in other words, the probability of the intersection of
multiple events. 
\\
\\
In Figure~\ref{figure:part2Tree}, by construction node $A$, through path $*\rightarrow B \rightarrow A$, 
has a size $\mathbb{P}(\Omega \cap B
\cap A)$, and node $B$ has size $\mathbb{P}(B)$. We can now endow the weight
of edge $B \rightarrow A$ according to the proportion of node $A$ (treated as a subset of $B$) 
out of $B$, or the probability of transition to node $A$ given that one has reached node $B$ 
from the root. This equals $\mathbb{P}(A|B)$. Such a definition is valid as the size of nodes 
$A, \overline{A}$ and $B$ satisfies $\mathbb{P}(A \cap B) + \mathbb{P}(\overline{A} \cap B)=\mathbb{P}(B)$.
\begin{figure}[h]
\centering
\begin{center}
\hspace{0cm}
\includegraphics[scale=0.42,clip]{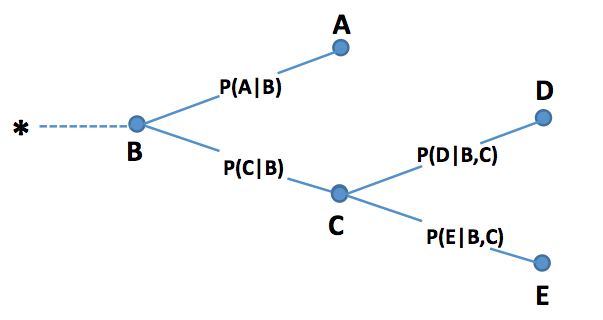}
\end{center}
\abovecaptionskip=-5pt
\caption{\it The tree diagram approach illustrated. } \label{figure:condProbDeft}
\end{figure}
\\
\\
Thus, in Figure~\ref{figure:condProbDeft}, the probability that one arrives at a node,
say $A$, along the path $\Omega \rightarrow B \rightarrow A$ is given by
\begin{equation}
\mathbb{P}(\Omega B A)=\mathbb{P}(\Omega \cap B \cap A)=\mathbb{P}(BA)=\mathbb{P}(B)\cdot \mathbb{P}(A|B),
\end{equation}
which is simply the product of edge weights along the path $\Omega \rightarrow
B \rightarrow A$ (the edge weight for $\Omega \rightarrow B$ is $\mathbb{P}(B)$). 
Same reasoning extends to any node in a tree. Thus we have provided a
tree-based interpretation of the turtleback diagram for conditional probability.
Such an algebraic system on the tree has the following two properties:
\begin{enumerate}
\setlength\itemsep{-0.02in}
\item The probability of arriving at any node equals the product of edge weights along the path.
\item The weight of an edge $H \rightarrow L$ has weight given by $\mathbb{P}(L | *, ..., H)$.
\end{enumerate}
This is exactly what a tree diagram would represent. The above properties extend readily to a series of events. 
For example, the probability of a series 
of events, $B \rightarrow C \rightarrow D$ can be computed as the probability of arriving at node 
$D$ along the tree path $\star \rightarrow B \rightarrow C \rightarrow D$ (c.f., Figure~\ref{figure:condProbDeft})
\begin{eqnarray}
\mathbb{P}(B \cap C \cap D) &=& \mathbb{P}(* \rightarrow B)\cdot \mathbb{P}(B \rightarrow C)\cdot \mathbb{P}(C \rightarrow D) \nonumber\\
&=& \mathbb{P}(B) \cdot \mathbb{P}(C | B) \cdot \mathbb{P}(D | B, C).
\end{eqnarray}
This approach applies even for non-sequential events, as one can artificially attach an order to the 
events according to the ``arrival" of relevant information. Thus, we have shown the semantic equivalence 
between the turtleback diagram and the tree diagram. Their difference is mainly on the visual representation, 
which matters as visual tools.
\\
\\
The tree diagram appears to be less intuitive than the turtleback diagram as there is no longer an 
association between the area of a region and its probability (one may use the thickness of an edge 
to indicate the probability, but that is less attractive too). However, 
the tree diagram seems to scale better to large problems.
\section{Case studies}
\label{section:caseStudy} We consider four examples in case study, including the {\it `Lung disease and smoking'} 
example, the {\it `History and war'} example, the {\it `Lucky draw'} example, and
{\it `the urn model'} example \cite{Rice1995}. As a matter of fact, very few students (about $10-15\%$) can do the `History and war'
example completely correctly in an in-class practice, after explaining to them the non-graph based concept of conditional probability.
That motivated us to adopt the graph-based approach. In the following, we provide the details of the examples.
\subsection{The lung disease and smoking example}
\label{section:lungSmokert}
\begin{figure}[h]
\centering
\begin{center}
\hspace{0cm}
\includegraphics[scale=0.42,clip]{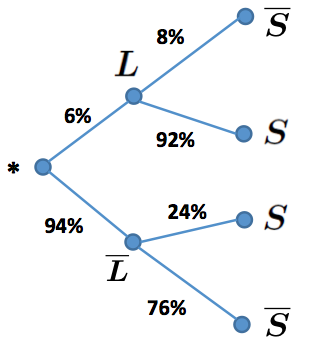}
\end{center}
\abovecaptionskip=-2pt
\caption{\it The tree diagram approach for the `Lung disease and smoking' example.
The letters $``L"$ and $``\overline{L}"$ stand for ``with lung disease" and ``without lung
disease", $``S"$ and $``\overline{S}"$ for ``smoking" and ``nonsmoking", respectively.
} \label{figure:LSHW}
\end{figure}
\noindent
With the tree diagram, the answer to (1) is the probability of reaching
node $S$ along the path $\star \rightarrow L\rightarrow S$, which is the product of
edge weights along this path and is calculated as $6\%
\cdot 92\%=0.0552$. The solution to (2) is the sum of products of edge weights along two
paths, $\star \rightarrow L\rightarrow S$ and $\star \rightarrow
\overline{L}\rightarrow S$, that is, $6\% \cdot 92\%+94\% \cdot 24\%=0.2808$, and (3) by the
ratio of the product of edge weights along path $\star \rightarrow
L\rightarrow$ over that over two paths, which is $0.0552/0.2808=0.1966$.
\subsection{The History and War example}
\label{section:historyWar} 
\begin{figure}[h]
\centering
\begin{center}
\hspace{0cm}
\hspace{-0.1in}\includegraphics[scale=0.32,clip]{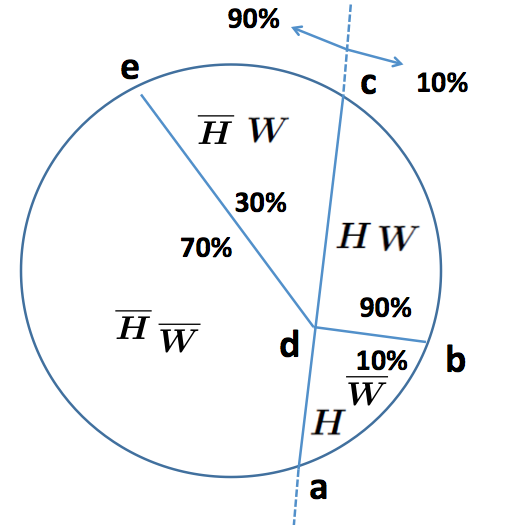}\hspace{0.2in}
\includegraphics[scale=0.42,clip]{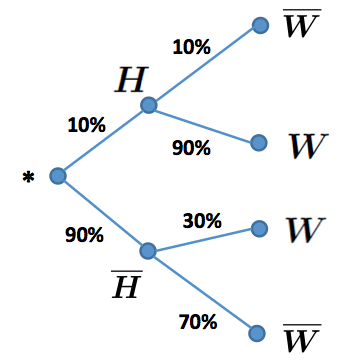}
\end{center}
\caption{\it Solving the `History and War' example with the turtleback diagram and tree diagram, 
respectively. The letters $``H"$ and $``\overline{H}"$ stand
for ``like movies related to history" and ``do not like movies related to history",
$``W"$ and $``\overline{W}"$ for ``like movies related to wars" and ``do not like movies
related to wars", respectively.
} \label{figure:HWgt}
\end{figure}

This example is artificially created so that it
has a similar problem structure as the {\it `Lung disease and smoking'} example.
It is described as follows.
\begin{quote}
``According to a market research about the preference of movies, $10\%$ of
the population like movies related to history. Of those who like movies related
to history, $90\%$ also like movies related to wars; of those who do not like
movies related to history, only $30\%$ like movies related to wars. Answer
the following questions.
\begin{enumerate}
\item[(a)] If a person is randomly selected in the population, what is
    the chance that she likes both movies related to wars and movies related
    to history?
\item[(b)] If a person is randomly selected in the population, what is
    the chance that she likes movies related to wars?
\item[(c)] If a person is randomly selected and is discovered to like
    movies related to wars, what is the chance that she likes movies
    related to history?"
\end{enumerate}
\end{quote}
We can construct a turtleback diagram as the left panel of Figure~\ref{figure:HWgt}.
One can quickly answer the questions as
follows. (a) is the area of region ``adba", which is given by $10\% \cdot
90\%=0.09$, (b) is the total area of region ``edbae", which is given by $10\%
\cdot 90\%+90\% \cdot 30\%=0.36$, and (c) is the ratio of (a) and (b) which is $0.09/0.36=0.25$.
\\
\\
Similarly, the right panel of Figure~\ref{figure:HWgt} is a tree diagram. One can answer the 
questions as follows. (a) is the product of edge weights along the path $\star \rightarrow H
\rightarrow W$, which is given by $10\%
\cdot 90\%=0.09$, (b) is the sum of the product of edge weights along two paths,
$\star \rightarrow H \rightarrow W$ and $\star \rightarrow \overline{H}
\rightarrow W$, which is given by $10\% \cdot  90\%+90\% \cdot 30\%=0.36$, and (c) is
the ratio of (a) and (b) which is $0.09/0.36=0.25$. 
\subsection{The lucky draw example}
\label{section:luckyDraw} The {\it lucky draw} example is taken from the popular
lucky draw game. This example is especially useful as many sampling without
replacement problems can be converted to this and solved easily. Here we take
a simplified version with the total number of tickets being $5$ and there is only
one prize ticket. The description is as follows.
\begin{quote}
``There are $5$ tickets in a box with one being the prize ticket. $5$ people
each randomly draws one ticket from the box without returning the drawn ticket
to the box. Is this a fair game (i.e., each draws the prize ticket with the same 
chance)?"
\end{quote}
\begin{figure}[h]
\centering
\begin{center}
\hspace{0cm}
\includegraphics[scale=0.36,clip]{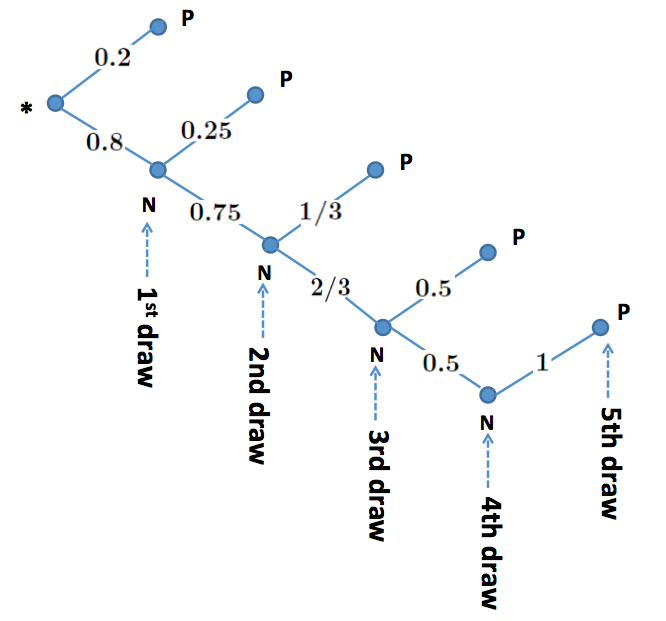}
\end{center}
\abovecaptionskip=-1pt
\caption{\it The tree diagram for the `Luck draw' game. The letters $``P"$ and
$``N"$ denote the prize ticket and non-prize ticket, respectively. 
} \label{figure:luckyDrawt}
\end{figure}
Figure~\ref{figure:luckyDrawt} depicts the process of ticket drawing. As here
our interest is the prize ticket, the tree branch that has already
seen the prize ticket will not grow further. Easily the probability of
getting the prize ticket at the first draw is $1/5$. Following
Figure~\ref{figure:luckyDrawt}, the probability of getting the prize ticket
at the second draw is the product of edge weights along the path $\star
\rightarrow N \rightarrow P$, which is $0.8 \cdot 0.25=0.2$. Similarly, the
probability of getting the prize ticket at the third draw is given by
$0.8 \cdot 0.75 \cdot 1/3=0.2$, and so on. 
\begin{figure}[h]
\centering
\begin{center}
\hspace{0cm}
\includegraphics[scale=0.4,clip]{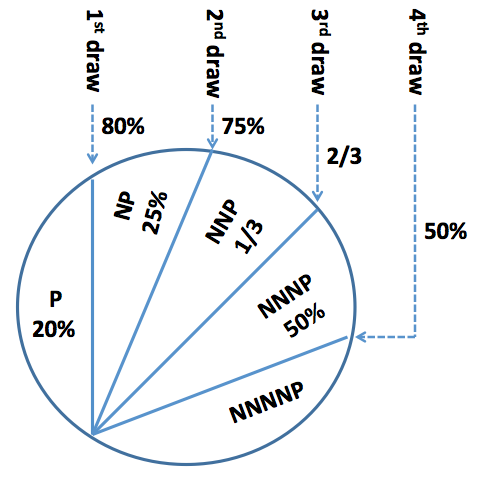}
\end{center}
\abovecaptionskip=-3pt
\caption{\it The turtleback diagram for the `Luck draw' game. The letters in the labels indicates status of each attempt, 
$``P"$ for prize and $``N"$ for a non-prize ticket. For example, ``NNP" means getting non-prize tickets for the first two draws and
the prize ticket at the third draw. The percentage next to the label indicates the probability of a prize at the last draw, conditional on 
the outcome of all preceding draws. For example, ``25\%" next to ``NP" means the conditional probability of getting a prize is 25\% in 
the second draw if the first draw is not a prize. Or in other words, that is the ratio of the area of the slice containing ``NP" to all slices after 
the first slice is taken away.   
} \label{figure:luckyDrawturtle}
\end{figure}
\noindent
\\
\\
Figure~\ref{figure:luckyDrawturtle} is the turtleback diagram for the `Luck draw' game. 
Easily the probability of getting the prize ticket at the first draw is the area of the region 
labelled as ``P", which is $0.2$. Following the figure, the probability of getting the prize ticket
at the second draw is the area of the region labelled as ``NP", which is $0.8 \cdot 25\%=0.2$. 
Similarly, the probability of getting the prize ticket at the third draw is given by
$0.8 \cdot 75\% \cdot 1/3=0.2$, and so on. 
\subsection{An urn model example}
\label{section:urnModel} This can be viewed as an extension of the lucky draw
problem in the sense that there are more than one prize tickets here. Note
that this example mainly serves to demonstrate that both the tree and the turtleback 
diagram could be used to solve problems of such a complexity (one can solve this problem quickly by distinguishing
the two green balls and apply result of the lucky draw game\footnote{Label
the two green balls as $G_1$ and $G_2$, respectively. Then the probability of
getting a green ball at each draw is simply that of getting $G_1$ or $G_2$.
Either $G_1$ or $G_2$ can be treated as the only prize ticket in the lucky
draw game thus the probability of getting either one is $1/5$, and so the
probability of getting a green ball at any draw is always $2/5$.}). Assume there are $2$ greens balls and $3$ red balls.
The problem is described as follows.
\begin{quote}
``There are $2$ green balls and $3$ red balls in an urn. One randomly picks
one ball for five times from the urn without returning. Will each draw have
the same chance of getting the green ball?"
\end{quote}
\begin{figure}[h]
\centering
\begin{center}
\hspace{0cm}
\includegraphics[scale=0.4,clip]{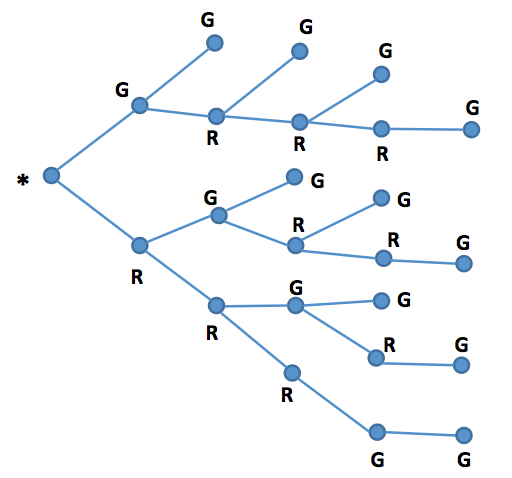}
\end{center}
\abovecaptionskip=-5pt
\caption{\it The tree-based approach for an urn model with $2$ green balls and $3$
red balls. We use the $\star$ to denote the root node, and ``R" and ``G" for red ball and green ball, respectively. As our
interest is the green balls, so tree branches that have seen $2$ green balls
will not grow any further.} \label{figure:2G3R}
\end{figure}
\begin{figure}[h]
\centering
\begin{center}
\hspace{0cm}
\includegraphics[scale=0.4,clip]{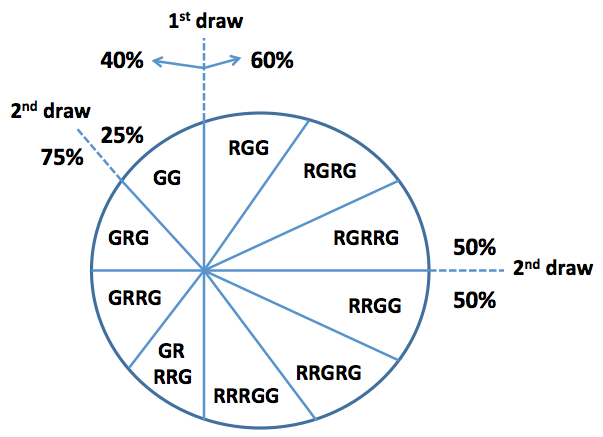}
\end{center}
\abovecaptionskip=-3pt
\caption{\it The turtleback diagram for an urn model with $2$ green balls and $3$
red balls. ``R" and ``G" are used to denote red ball and green ball, respectively. As our
interest is the green balls, so tree branches that have seen $2$ green balls will not 
grow any further. Each letter, ``G" or ``R", indicates the outcome of a particular draw. 
For example, a ``RGRG" indicates that the first draw gets a red ball, the second draw 
a green ball, the third a red and the fourth a green ball.} \label{figure:2G3Rturtle}
\end{figure}
Figure~\ref{figure:2G3R} is the tree diagram for the urn model.  
We are not going to calculate the probability of getting a green ball for
each draw, instead we only do it for the third draw. The probability of
getting a green ball at the third draw is give by the sum of the product of
edge weights along three paths
\begin{equation*}
\star \rightarrow G \rightarrow R
\rightarrow G, ~\star \rightarrow R \rightarrow R \rightarrow G, ~\star
\rightarrow R \rightarrow G \rightarrow G,
\end{equation*}
which is
$(2/5)(3/4)(1/3)+(3/5)(1/2)(2/3)+(3/5)(1/2)(1/3)=2/5$. One can similarly
calculate that the probability of getting a green ball at other draws all
equal to $2/5$.
\\
\\
Figure~\ref{figure:2G3Rturtle} is the turtleback diagram for the urn model. To calculate the probability that the 
third draw gets a green ball, we simply sum up
the area of all regions with a label such that the third letter is ``G". That is, the total area of regions labelled as 
\begin{equation*}
``RGG", ~``GRG", ~``RRGG", ~``RRGRG", 
\end{equation*}
which is
\begin{equation*}
\frac{3}{5} \cdot \frac{1}{2}  \cdot \frac{1}{3} + \frac{2}{5} \cdot \frac{3}{4} \cdot \frac{1}{3}  + \frac{3}{5} \cdot \frac{1}{2}  \cdot \frac{2}{3} \cdot \frac{1}{2}  + \frac{3}{5} \cdot \frac{1}{2} \cdot \frac{2}{3} \cdot \frac{1}{2} =0.4.
\end{equation*}
The calculation seems a little tedious, but conceptually very simple, as long as one could follow the way the regions are partitioned.
\small
\vspace{5pt}
\begin{table}[h]
\begin{center}
\begin{tabular}{c|c|c|c}
\textbf{Course} &\textbf{\# students} & \textbf{Class size}  & \textbf{Institute} \\
\hline
\rule{0pt}{2.5ex} STAT235 & 128 & 40-60 &UMKC\\
\hline
\rule{0pt}{2.5ex}MTH231     &  25& 10-20 & UMassD\\
\hline
\rule{0pt}{2.5ex}MTH331 &  72 & 35-45 & UMassD\\
\hline
\end{tabular}
\caption{\it Students involved in the case studies.}
\label{table:students}
\end{center}
\end{table}
\normalsize
\section{Empirical data}
\label{section:caseStat}
We carried out case studies on over $200$ students. This includes students in the 
elementary statistics class, STAT235 (non-calculus based), at University of Missouri Kansas City (UMKC) 
during 2012-2013, and students from elementary statistics, MTH231, and elementary probability, MTH331, 
classes at University of Massachusetts Dartmouth (UMassD) during 2015-2017. These three courses had 
a fairly different student population. For STAT235, about 30\% from engineering, 30\% from business, and 
the rest from such diverse majors as biology, chemistry, psychology, political sciences, education etc.
For MTH231, about 80\% are from mathematics or data science, and the rest from majors such as computer 
science, electrical engineering, criminal justice etc. For MTH331, about 75\% from computer science or electrical 
engineering, 20\% from mathematics or data science, and the rest from other engineering majors or economics, 
physics etc. Table~\ref{table:students} gives a summary of students involved in the case studies.
\\
\\
The study is carried out as follows. First we explain to students the concept of conditional probability with 
a non-graph based approach. Then we continue with two exercises. In the first exercise, we explain to 
students the `Lung disease and smoking' example, with both the turtleback and the tree diagram, and 
have students solve the `History and war' problem, or vice versa (for different classes we were teaching). 
In another exercise, we explain the `Lucky draw' example, and have students solve the `Urn model' problem,
or vice versa. Due to time constraints on the course schedule, we did not ask students to solve problems 
using a particular technique followed by its discussion. Rather we discussed both the turtleback and the tree 
diagrams, and let students choose one of them for problem solving. Table~\ref{table:expStatBreakdown} is a 
breakdown of the number of students involved.
\\
\small
\begin{table}[h]
\begin{center}
\begin{tabular}{c|c|c|c|c}
 \textbf{Course}		             &\textbf{Lung disease and}    & \textbf{War and history}   & \textbf{Lucky}       & \textbf{Urn} \\
                                             &   \textbf{smoking}                & \textbf{movie} & \textbf{draw}  & \textbf{model}\\
\hline
\rule{0pt}{2.5ex} STAT235 & 66 &62  &62  & 66 \\
\hline
\rule{0pt}{2.5ex} MTH231 & 14 & 11 &14  &11  \\
\hline
\rule{0pt}{2.5ex} MTH331 &  37 &35  &35  & 37 \\
\hline
\rule{0pt}{2.5ex}Total & 117 & 108 & 111 & 114 \\
\hline
\end{tabular}
\caption{\it Number of students involved in the empirical study breakdown by course and problem.}
\label{table:expStatBreakdown}
\end{center}
\end{table}
\small
\begin{table}[h]
\begin{center}
\begin{tabular}{c|c|c|c|c}
 		&\textbf{Neither} & \textbf{Either one} & \textbf{Prefer}       & \textbf{Prefer} \\
  \textbf{Question}                     &\textbf{helpful}  & \textbf{helpful}        & \textbf{Turtleback} & \textbf{Tree} \\
\hline
\rule{0pt}{2.5ex} Lung disease and smoking & 13.7\% & 86.3\% & 53.0\% & 33.3\% \\
\hline
\rule{0pt}{2.5ex} War and history movies     &  11.1\%& 88.9\% & 54.6\% & 34.3\% \\
\hline
\rule{0pt}{2.5ex} Lucky draw &  17.1\% & 82.9\% & 34.2\% & 48.6\% \\
\hline
\rule{0pt}{2.5ex} Urn model &21.9\% &78.1\% & 31.6\% & 46.5\% \\
\hline
\end{tabular}
\caption{\it Data collected in the case studies on whether graphs help understand the concept of conditional probability, and the 
preference between the turtleback diagram and the tree diagram.}
\label{table:expStat}
\end{center}
\end{table}
\normalsize
\\We collect two types of data from the case studies, one on students' preference between graph and non-graph based 
approach, and the other on students' preference between the turtleback and the tree diagram. Here, except for the case
of non-graph based approach, by preference we mean the students actually used the technique for problem solving, and 
nearly in all such cases they could apply it correctly in solving the assigned problem; so we use this as measurement of 
learning outcome (with an understanding that further experiments may be needed to validate this). The results 
are reported in Table~\ref{table:expStat}. The data collected are quite encouraging. About 78-88\% students found a 
graph tool helpful. For the `Lucky draw' and the `Urn model', fewer students found it helpful. This is possibly because 
these two problems appear to be harder to students: even a graphical tool may not help them much. 
Further experiments are needed to validate or understand this.
\\
\\
In terms of a preference for which graphical tool, the results show an interesting pattern. For the `Lung disease and smoking' 
and the `War and history' example, more students prefer the turtleback diagram to the tree diagram, around 53-54\% vs 
33-34\%. The `Lucky draw' and the `Urn model' examples exhibit an opposite pattern, more students prefer the tree 
diagram to the turtleback diagram, around 46-48\% vs 31-34\%\footnote{Since in all cases, the sample size is large 
enough and the difference between contrast groups is significant, we did not carry out a hypothesis testing using the 
reported data.}. This is probably due to the fact that, in the first two examples, the sample spaces and events involve 
populations in the usual sense, while the last two examples involve sequential decisions, for which a tree structure that 
represents the decision dichotomy may be more natural (although in such cases, the concept of conditional probability 
is not as natural as that in the turtleback diagram). Further experiments are needed to confirm this. The advantage of 
the turtleback diagram over the tree diagram appears to decrease as the problem becomes harder, but this is not a 
serious problem for beginning students as {\it those who most need help from a graphical representation are just those 
who could not solve simple problems}. Moreover, we do not expect one single graphical tool can help solve all the problems, 
rather different people may use different tools for a particular problem.
\section{Potential research questions}
Many instances of conditional probability occur in sampling without replacement. Tarr and Jones \cite{TarrJones1997} describe a framework for assessing middle school students' thinking in conditional probability and independence, which is elaborated in \cite{TarrLannin2005}. This framework is a levels model, with 4 levels---\textit{Subjective}, \textit{Transitional}, \textit{Informal Quantitative}, and \textit{Numerical}---subject to all the difficulties such a model has as students transition from one level to another. 
\\
\\
\textbf{Research Question 1}: Are turtleback diagrams, as compared to tree diagrams, helpful to students, at any or all of the Tarr-Jones framework levels, in understanding conditional probability. If so, how can we measure and assess the comparative utility of turtleback diagrams compared to tree diagrams?
\\
\\
\textbf{Research Question 2}: Related to Research Question 1, specifically, how helpful are turtleback diagrams in helping students understand conditional probability in the context of sampling without replacement?
\\
\\
Conditional probability is increasingly being introduced into middle school in the United States. The Conference Board of the Mathematical Sciences \cite{ConferenceBoard2001} stated:
\begin{quote}
Of all the mathematical topics now appearing in middle grades curricula, teachers are least prepared to teach statistics and probability. Many prospective teachers have not encountered the fundamental ideas of modern statistics in their own K-12 mathematics courses...Even those who have had a statistics course probably have not seen material appropriate for inclusion in middle grades curricula. (p. 114)
\end{quote}
\textbf{Research Question 3}: Are turtleback diagrams helpful to middle school teachers of probability and statistics in (a) enhancing their own understanding of conditional probability and (b) assisting them to better teach conditional probability? If so, how and to what extent?

\section{Conclusions}
\label{section:conclusions} Motivated by difficulties encountered by many
undergraduate students new to statistics, we re-examined the definition and representation of conditional probability, and presented a Venn-diagram like approach: the turtleback diagram. We discussed our graphical tool in the context of other graphical models for conditional probability, and carried out case studies on over $200$ students of elementary statistics or probability classes. 
Our case study results are encouraging and the graph-based approaches could potentially lead to significant improvements in both the students' understanding of conditional probability and problem solving. 
While the existing tree diagram is preferred to the turtleback diagram on problems 
that involve a sequential decision, the turtleback diagram is considered more helpful 
in settings where the underlying population resembles the usual human population; it is 
exactly in such situations that weaker students are more likely to need help. 
Though the turtleback diagram appears very different from the tree diagram, we are 
able to unify them and show their equivalence in terms of semantics. 
\\
\\
Our discussion suggests
a simple framework for visualizing abstract concepts, that is, a suitable graph representation of the abstract 
concept followed by a simple post-processing in the visual-brain system. A good visualization idea needs to 
balance both. We are able to use such a framework to interpret the difficulty encountered by the tree diagram, 
and aid our development of the turtleback diagram. Further studies are expected to validate or to adopt such 
a framework to general visualization tasks. Given the increasingly important role played by data visualization 
in data science and exploratory data analysis \cite{Tukey1977, TufteVisual1983,Cleveland1993,Yau2011}, it 
would be worthwhile to give a few remarks here comparing the graph representation of abstract concepts and 
data visualization. These two concepts are different yet closely related. Graphical representation aims to 
understand an abstract (or complicated) concept by representing elements of the concept with a graph, while 
data visualization seeks to understand the data or the information behind by displaying aspects (i.e., descriptive 
statistics) of the data. In terms of implementation, as both aim to help understanding or reasoning, the used 
graphical objects need to be simple (though simple in different ways in the two cases). In data visualization, 
the graphical objects need to be simple so that people can quickly grasp the information conveyed or to understand 
the concept behind without resorting to paper and pencil; in graphical representation of concepts, the objects 
need to be conceptually simple and easy to manipulate for applications of the concepts. 
\\
\\
Our case studies suggest that it is worthwhile to introduce such graphical tools to students whose success 
would seem to depend on them. We hope that this will benefit our statistics colleagues who are teaching 
elementary statistics and students who are struggling with the concept of conditional probability and its 
application to problem solving. The potential savings in time can be huge. As a conservative estimate, assume
each year there are about $1.5$ million bachelor's degrees awarded in US (about $1.67$ million awarded 
in 2009). Assume there are about $200,000$ of them have taken an elementary statistics class, and about 
$10\%$ of them need help and succeed with our proposed approach, and further assume an average class 
size of $40$. If each instructor saves $2$ hours of time in each elementary statistics class and each student 
who benefits from our approach saves $1$ hour, then the estimated total amount of time saved is at least 
$30,000$ hours {\it per} year in the U.S. alone.
\section*{Acknowledgement} 
\addcontentsline{toc}{section}{Acknowledgement}
The authors are grateful to Professor Yong Zeng at UMKC for kindly pointing to 
the `Lung disease and smoking' example, and for encouragement and support on some of the case studies.

\end{document}